\documentclass[journal,twoside,web]{ieeecolor}
\usepackage{tdei}
\usepackage{amsmath,amsfonts}
\usepackage{algorithmic}
\usepackage{algorithm}
\usepackage{array}
\usepackage[caption=false,font=normalsize,labelfont=sf,textfont=sf]{subfig}
\usepackage{textcomp}
\usepackage{stfloats}
\usepackage{url}
\usepackage{verbatim}
\usepackage{graphicx}
\usepackage{cite}
\usepackage{color}
\begin{document}

\title{Free Volume Model Analysis of the O\(_{2}^{-}\) Ion Mobility in Dense Ar Gas}

\author{Armando Francesco Borghesani and Fr\'ed\'eric Aitken %,~\IEEEmembership{Staff,~IEEE,}
        % <-this % stops a space 
\thanks{This work benefited from the support of the project ZEROUATE under Grant ANR-19-CE24-0013 operated by the French National Research Agency (ANR).}
\thanks{
Armando Francesco Borghesani (corresponding author) is with CNISM Unit, Department of Physics \& Astronomy, Universit\`a degli Studi, Padua, Italy (e-mail: armandofrancesco.borghesani@unipd.it)  %(Corresponding author: A. F. Borghesani)
.}% <-this % stops a space
\thanks{Fr\'ed\'eric Aitken is with G2ELab, C.N.R.S., University of Grenoble Alpes, Grenoble, France (e-mail: frederic.aitken@g2elab.grenoble-inp.fr)} %
\thanks{Both authors are co-first authors.}
%\thanks{Manuscript received xx, yy, zzzz; revised xy, yx,www5.}}
} %
\markboth{\journalname,~Vol.~xx, No.~y, zzz~wwww}%
{Borghesani and Aitken: Free Volume Model Analysis of the O\(_{2}^{-}\) Ion Mobility in Dense Ar Gas}

%\IEEEpubid{0000--0000/00\$00.00~\copyright~zzzz IEEE}

\maketitle

\begin{abstract}
  We report the  results of the experiment aimed at measuring the mobility of %negatively charged
   O\(_2^-\) ions in dense argon gas in the temperature range \(180\,\mbox{K}\leq 
  T \leq 300\,\)K. We show that an adequate theoretical description of the data is obtained by using the thermodynamic Free Volume (FV) model, originally developed to describe the electron bubble mobility in superfluid helium and successfully exploited for describing the \(O_2^-\) mobility in near critical neon gas.  The model goal is to thermodynamically predict the free space available for ion motion. By implementing the FV model with the Millikan-Cunningham (MC) slip correction factor, we can describe the ion mobility in the crossover region bridging the dilute gas kinetic regime to the high-density hydrodynamic regime of ion transport.
These results confirm the validity of the model and the universality of some of its features.
\end{abstract}

\IEEEpubidadjcol
\section{Introduction}
\IEEEPARstart{T}{he} detailed knowledge of the transport properties of ions in dense gas- and liquid dielectrics is the cornerstone for understanding, designing, and harnessing processes in several natural as well as applicative fields. 
Control over biological processes, chemical synthesis protocols, electrical discharge equipment facilities, among many others, requires that the behavior of ions drifting under the action of electrical fields is known and/or predictable in advance. Also, the behavior of high-energy detectors~\cite{lopez2005} and low-temperature plasma processes is affected by the transport behavior of ions~\cite{Bruggeman2016b}. Moreover, the profound comprehension of the physico-chemical mechanisms of the processes occurring in the atmosphere, so important to ascertain several aspects of climate change, is based on the knowledge of the ion transport regimes~\cite{mason2001}. 
Additionally, last but not least, the investigation on how ions drift in a dense disordered dielectric medium can shed light on the fundamental ion-neutral interaction mechanisms and provides pieces of information on the ion-neutral interaction potentials~\cite{viehland1976,Viehland1995}.

Typically, two different regimes for ion transport are considered. In dilute gases, the ion mobility \(\mu\) is mainly determined by binary collisions with the host gas atoms and the classical kinetic theory gives an accurate description of \(\mu\) as a function of electric field and temperature provided that the ion-atom scattering cross section is known and the appropriate collision integrals can be computed.  Conversely, mobility data can be used to invert the equations of the classical kinetic theory to determine the interaction potential~\cite{viehland1975,viehland1975b,maitland}. 
 If the neutral-ion interaction can be modeled as a hard-sphere interaction with hard-core radius \(R_0\), the density-normalized mobility \(\mu N\) of the ions in thermal equilibrium with the gas, as is the case of the present experiment, is obtained as
 \begin{equation}\label{eq:munkinetic}
 \mu N= \frac{3}{2\sigma}
  \left(\frac{\pi}{2m_r k_\mathrm{B} T}\right)^{1/2}
\end{equation}
in which \(\sigma = \pi R_0^2\) is the scattering cross section, \(N\) is the gas number density, \(T\) is the temperature, \(k_\mathrm{B}\) is the Boltzmann constant, and \(m_r\) is the ion-neutral reduced mass.

At the other boundary of the thermodynamic state of the medium, ions drift in a liquid and their thermal mobility is described by the Stokes hydrodynamic formula
\begin{equation}\label{eq:purestokes}
\mu N = \frac{eN}{6\pi \eta R}
\end{equation}
in which \(e\) is the charge of the (singly charged) ion, \(\eta\) is the gas viscosity, and \(R\) is the so-called hydrodynamic (or, effective) radius, which is assumed to be a constant for a given ion-neutral pair~\cite{byron,mason1988}.

However, this experiment, as several more similar ones~\cite{Borghesani1993,Borghesani1995,Borghesani1997}, is carried out in a dense gas, for which a full-fledged theory for ion transport encompassing the crossover region between the dilute gas limit and the hydrodynamic regime is absent. Moreover, negative ions are not extensively investigated as the positive ones because of the difficulty of their production. Whereas positively charged ions are produced by direct ionization, the formation of a negative ion goes through several steps, which may not necessarily occur in the correct sequence. Firstly, low-energy electrons must be captured by electronegative molecular impurities. Then, the vibrationally excited transient anions thus formed quickly decay by autodetachment but a fraction of them can be stabilized by collisions with a third body, typically a host gas atom that carries away the excess energy~\cite{Bradbury1933,christophourou1984a}. Only this fraction of ions is sufficiently long-lived to allow the experimenters to measure their mobility. Only a small number of stable ions is made available because this negative ion production mechanism is not efficient and strongly depends on the environment. The delicate balance of short-range repulsive exchange forces acting between the excess electron in the ion and the electronic clouds of the surrounding atoms, and the long-range polarization interaction leads to the birth of a state that cannot adiabatically be obtained by simply adding an ion to the medium. Therefore, negative ions are endowed with a structure more complicated than that of cations. The ion is localized in an empty cavity surrounded by a solvation shell produced by electrostriction. The properties of the structure depend on the atomic polarizability of the gas and on its thermodynamic state~\cite{Volykhin1995,Khrapak1995,Volykhin1999,Schmidt1999}.

Over the years, we have been carrying out several experimental investigations on the mobility of  O\(_2^-\) ions in dense noble gases in wide temperature and density ranges~\cite{Borghesani1993,Borghesani1995,Borghesani1997,Borghesani1999}. 
Oxygen is the most ubiquitous and important attaching species for evident reasons and is readily available as an impurity even in the best purified gas. Being the same ionic species in different gases its behavior sheds light on gas specific features. 

The choice to carry out measurements in a supercritical or near-critical gas is dictated by the fact that the gas density can be varied at will by changing pressure or temperature.
By so doing, also the Knudsen number, \(K_n= \ell (N,T)/R\), which is the ratio of the  mean free path \(\ell\) of the neutrals to the ion effective radius, can be varied in quite a range, thereby bridging the \(K_n\ll 1\) hydrodynamic regime to the \(K_n \ge 1\) kinetic regime.

As pointed out several times ~\cite{Borghesani1993,Borghesani1995,Borghesani1997,Borghesani1999,Borghesani2019,Borghesani2020}, in the broad temperature and density ranges we have explored, neither the kinetic- nor the hydrodynamic description of the experimental data are adequate. However, a thermodynamic model (the FV model) has been recently developed that 
aims at computing the \(N\) and \(T\) dependence of the free volume available for the ion motion via an equation of state (EOS)~\cite{Aitken2011,Aitken2011a,Aitken2017}. The size of the free volume, which is of spherical shape based on symmetry arguments, is assumed to be the effective ion radius to be inserted in the Stokes formula %in 
of the hydrodynamic regime. The extension of the model prediction to smaller densities to describe the crossover region towards the kinetic regime is obtained by adopting a suitably modified form of the Millikan-Cunningham slip correction factor~\cite{Cunningham1910,Millikan1910,Tyndall1938}.

Despite being originally developed to describe the mobility of electron bubbles in superfluid Helium, the model has proven very effective to also describe the mobility of any kind of ions in several liquids and gases. We have successfully adopted this FV model to describe the mobility of O\(_2^-\) ions in dense (supercritical and near-critical) Neon gas~\cite{Borghesani2019,Borghesani2020} in extremely broad \(N\) and \(T\) ranges. 

Owing to the efficacy of the FV model and to test its validity for other systems we have carried out measurements of the O\(_2^-\) mobility in Argon gas in the temperature range  \(180\,\mbox{K}\le T\le 300\,\)K. We report here the new data and the result of their analysis with the FV model.

\section{Experimental Details\label{sect:expdet}}
\begin{figure}[h!]
\centering
\includegraphics[width=\columnwidth]{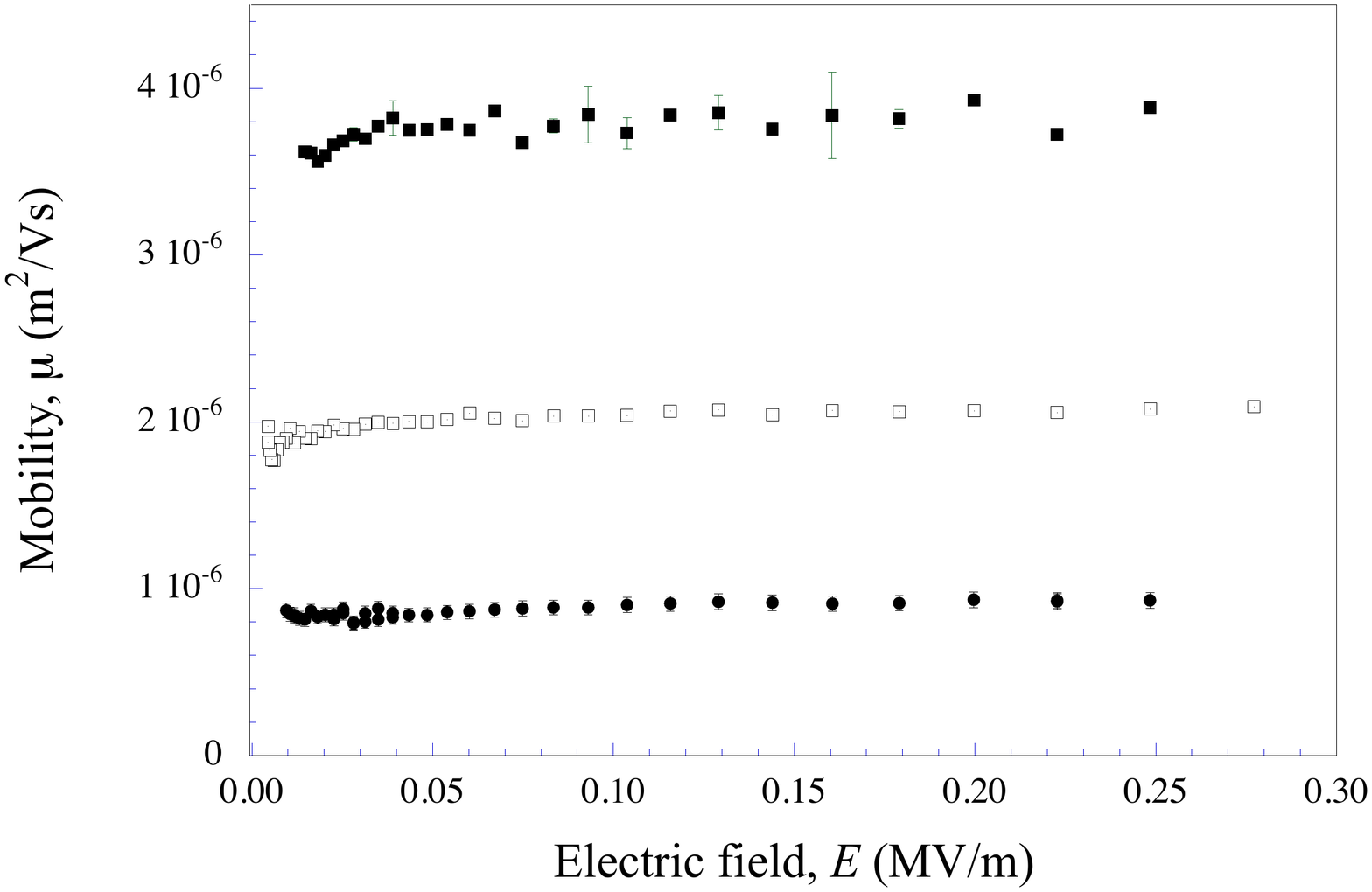}
\caption{\small Electric field dependence of \(\mu\) for \(T=180\,\)K and  \(N\,(10^{26}\,\mbox{m}^{-3})=10.6,\, 19.9,\, \mbox{and }50.6\) (from top).\label{fig:MuvsEexp6AKNT180K}}
\end{figure}

\begin{figure}[h!]
\centering
\includegraphics[width=\columnwidth]{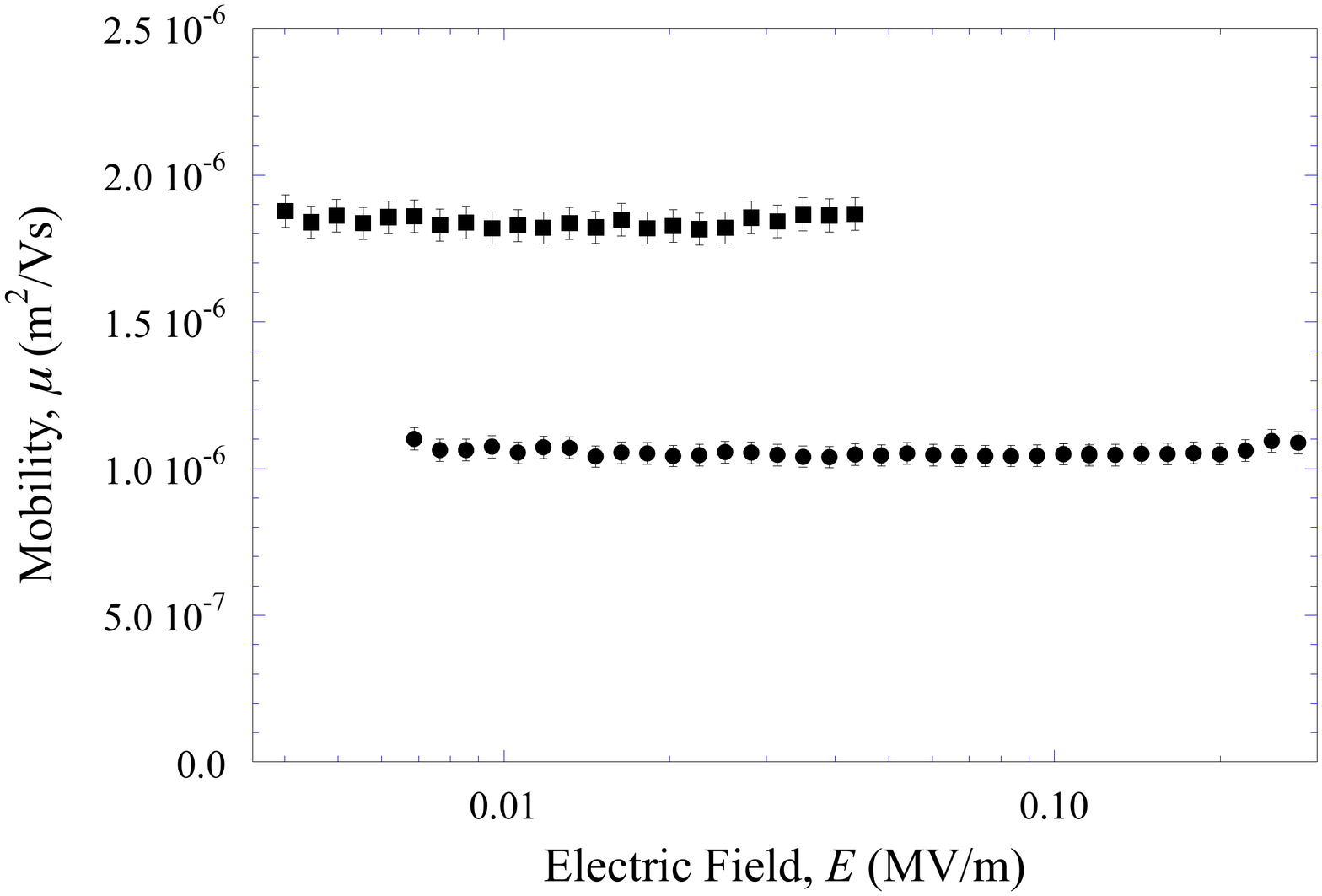}
\caption{\small Electric field dependence of \(\mu\) for \(T=260\,\)K and \(N\,(10^{26}\,\mbox{m}^{-3})=13.9\, \mbox{and }26.0\) (from top).\label{fig:O2-ArgonT260KN26Serie10Ae10F}}
\end{figure}

The experiment has thoroughly been described elsewhere~\cite{Borghesani1993}. 
  We briefly recall here its main features. We used the pulsed Townsend photo injection technique. The experimental cell can be pressurized up to \(\approx 8\,\)MPa and its temperature regulated within \(\pm 0.01\,\)K in the range \(25\,\mbox{K}\le T\le 340\,\)K. O\(_2^-\) are produced by resonant electron attachment to O\(_2\) molecular impurities in a concentration of a few p.p.m.~\cite{Borghesani2019}. Pressure readings are accurate within \(\pm 2\,\)kPa. The gas density is computed by means of an accurate equation of state~\cite{wagner}.
A d.c., 
high-voltage generator energizes the drift capacitor up to a maximum voltage of \(\approx 3\,\)kV. The drift distance is \(d\approx 1\,\)cm. The drift signal is passively integrated to improve the signal-to-noise ratio and the drift time is obtained by the analysis of the signal waveforms~\cite{Borghesani1990a}. 
 The explored range of reduced electric field is limited to \(E/N\le 40\,\)mTd (\(1\,\)mTd=\(10^{-24}\,\)V\(\,\)m\(^2\)). Thus, the ions always are in thermal equilibrium with the gas and their mobility \(\mu\) turns out to be field independent even at the lowest \(N\), as can be realized by inspecting Fig.~\ref{fig:MuvsEexp6AKNT180K} and. Fig.~\ref{fig:O2-ArgonT260KN26Serie10Ae10F} 

 We carried out measurements along several isotherms between \(T\approx 180\,\)K and \(T\approx 300\,\)K. The explored density range spans from \(N\approx 1\times 10^{26}\,\)m\(^{-3}\) up to \(N\approx 50\times 10^{26}\,\)m\(^{-3}\), well in the crossover region from the kinetic- to the hydrodynamic regime.

\section{Experimental Results}\label{sect:expres}
\begin{figure}[th!]
\centering\includegraphics[width=\columnwidth]{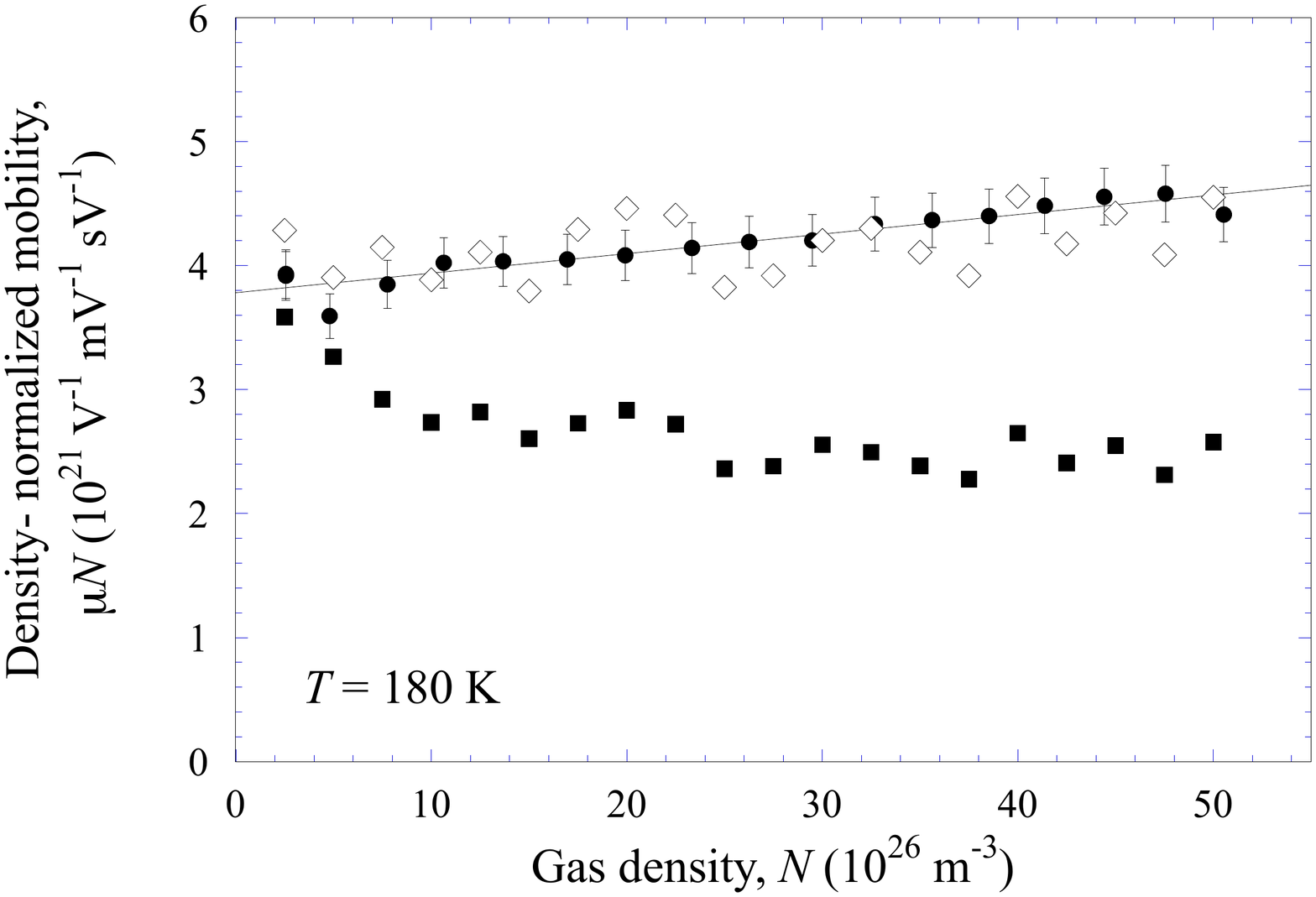}
\caption{\small Density dependence of \(\mu N\) in Argon gas for \(T=180\,\)K. Closed points: experiment.  Closed squares: Molecular Dynamics (MD) results. The uncertainty on the MD results is \(\approx 10\,\%\).  Open diamonds: MD results weighted by the the numbers of nearest neighbors. Solid line: linear fit to the experimental data. \label{fig:O2ArmuNT180&nearestneighbors}}
\end{figure}

\begin{figure}[t!]
\centering\includegraphics[width=\columnwidth]{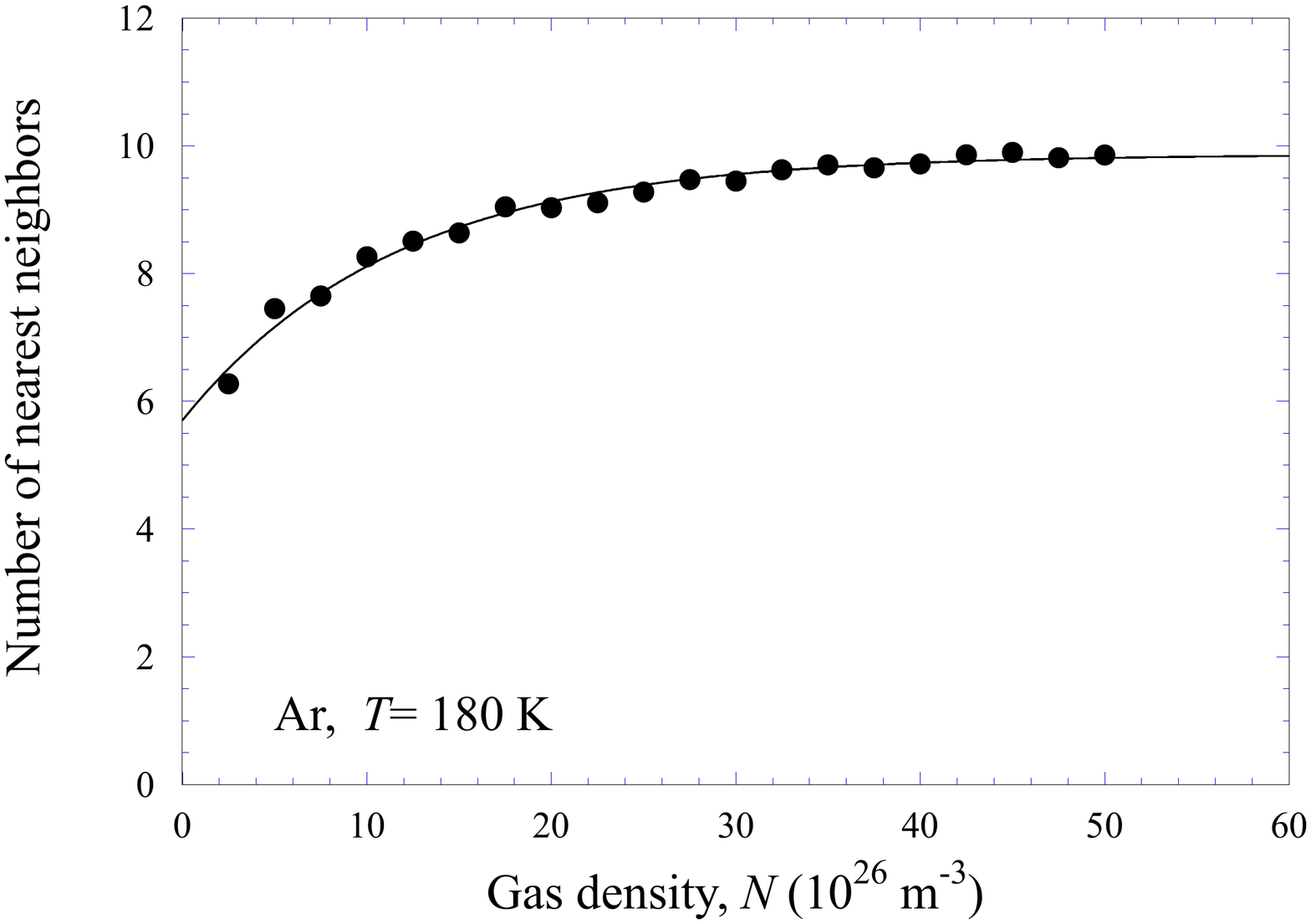}
\caption{\small Evolution with \(N\) of the number of the ion nearest neighbors  provided by MD simulations at \(T=180\,\)K. Solid line: only an eye guideline.\label{fig:nearestneighborsvsNT180K}}
\end{figure}

As previously put into evidence for Argon, Helium, and Neon~\cite{Borghesani2019,Borghesani2020}, the (zero-field) density-normalized mobility \(\mu N\) shows a weak, though evident, density dependence which can neither be described by the classical kinetic theory nor by the hydrodynamic Stokes formula. This observation is confirmed by these new measurements. We report in Fig.~\ref{fig:O2ArmuNT180&nearestneighbors}
the measured values of \(\mu N\) as a function of the density \(N\) for \(T=180\,\)K as closed points with error bars. The solid line is only a linear fit to the data.
To ascertain if this density dependence might stem from the type of interaction potential, we also carried out Molecular Dynamics (MD) simulations  as done elsewhere~\cite{Borghesani2018}. The MD results, shown as closed squares in Fig.~\ref{fig:O2ArmuNT180&nearestneighbors},
 have a completely different density dependence than the experimental data. However, the MD results can be reconciled with the experimental data (open diamonds in Fig.~\ref{fig:O2ArmuNT180&nearestneighbors}) if they are weighted by the number of nearest neighbors provided by the MD simulations shown in Fig.~\ref{fig:nearestneighborsvsNT180K}.

This observation clearly indicates that the ion mobility is strongly affected by the environment and that a description of the mobility in the crossover density region 
must be able to account for the structure of the ion environment. 
The goal of the FV model is just to provide a thermodynamic approach to description of the structure of the ion-medium complex.

\section{The free volume model\label{sect:FVM}}
The free volume model has successfully been adopted to rationalize both electron bubble-~\cite{Aitken2011,Aitken2016,Aitken2017} and positive ion mobility~\cite{Aitken2015} in liquid helium, but also O\(_2^-\) ion  mobility in low-density Helium~\cite{Aitken2011a} and in dense gaseous Neon~\cite{Borghesani2019,Borghesani2020}.
The main reasons for the success of the FV %M 
model  in different systems are that it computes in a thermodynamical way the effective ion radius, therefore accounting for the structure surrounding the ion itself, and that it exploits a modified form of the MC slip correction factor to bridge the two limiting transport regimes.

The cornerstone of the FV %M 
model is the concept of free volume \(V_f\), i.e., the volume available for the ionic motion through the medium. Ions are the solute species in the gas that acts as a solvent. The free volume is given by \(V_f=V-b\), in which \(V\) is the macroscopic volume occupied by the gas and \(b\) the covolume of its atoms.  As the ion number produced in the experiment is extremely low (their concentration is \(\le 10^{-14}\)), their covolume can safely be neglected.

The free volume per particle \(V_s=V_f/N\) is described by a van der Waals-like EOS
\begin{equation}
\label{eq:vs}
V_s=\frac{V_f}{N}=\frac{k_\mathrm{B}T}{P+\Pi}
\end{equation}
\(P\equiv P(N,T)\) is the hydrostatic pressure, which incorporates the attractive potential energy contributions among neutrals, and \(\Pi\) is the internal pressure that accounts for the excess attractive potential energy contributions in the systems, i.e., those due by the long-range polarization interaction between ions and neutrals.
The simplest analytical form for the internal pressure is \begin{equation}
\label{eq:PAI}\Pi = \alpha N^2
\end{equation}
in which \(\alpha\) is a constant that determines the density at which the compressibility is maximum and that has been proved to be system-independent according to the analysis of several ion-gas- and ion-liquids systems~\cite{Borghesani2019,Aitken2011a,Aitken2016,Aitken2015}.

 Owing to the spherical symmetry of the ion-neutral interaction, \(V_s\) is assumed to have spherical shape and its radius is considered to be the effective hydrodynamics radius \(R\).  As we expect that the size of the solvation shell depends on the ion-gas system compressibility, we must seek an analytic form relating \(R\) to \(V_s\) that accounts for this effect by enforcing agreement between the theoretical prediction and the experimental data.

To extend the range of applicability of the Stokes formula to crossover- and to lower density regions, the pure hydrodynamic Stokes formula is modified by the introduction of the empirical MC slip correction factor \(\phi(K_n,T)\), thereby leading to the following expression for the mobility
\begin{equation}
\label{eq:stokmod}
\mu N = \frac{eN}{6\pi\eta R}\left\{1+\phi\left[K_n(N,T)\right]
\right\}
\end{equation}
The behavior of \(\phi\) is such that it must vanish at very small Knudsen number, i.e., at high density, where the Stokes formula adequately describes the mobility, and that it must give (\ref{eq:munkinetic}) for large Knudsen number, i.e., at low enough density.
The Knudsen number is evaluated by using the kinetic theory expression for the mean free path of the neutrals~\cite{Reif}
\begin{equation}
\label{eq:elleta}\ell(N,T) = \frac{3\eta}{N\sqrt{8m_rk_\mathrm{B}T/\pi}}
\end{equation}
As the experiment is carried out at temperature quite higher than the critical one, we can use a version for the hydrodynamic radius with a much simpler temperature dependence than that used for neon gas close to the critical temperature~\cite{Borghesani2019,Borghesani2020}. Thus, we assume 
\begin{equation}
\label{eq:RsuR0}
\frac{R}{R_0} = 1+ \frac{\left(V_0/V_s\right)^{\epsilon_1}}{1+\gamma \left(V_0/V_s\right)}
\end{equation}
in which the parameters \(R_0\), i.e., the hard-sphere radius of the ion-medium interaction, \(\gamma\), \(\epsilon_1\), and the free volume of a suitable reference state, \(V_0=\delta k_\mathrm{B}T\), have all to be adjusted to give the best agreement with the experimental mobility data. 
We found \(R_0=0.555\,\)nm, \(\epsilon_1=12\), and \(\gamma=2\).
Moreover, we found that \(\alpha=0.8937\,\)MPa\(\cdot\)nm\(^6\) and \(\delta =0.15\,\)m\(^3\)/J. It is worth noting that the present values of \(\alpha\) and \(\delta\) are the same that give the best agreement in all investigated systems~\cite{Aitken2011,Aitken2011a,Aitken2015,Aitken2016,Borghesani2019,Borghesani2020} and also in several cation-liquids systems not yet published. Thus, the parameters \(\alpha\) and \(\delta\) appear to be universal. We believe that this universality arises as a consequence of adopting a van der Waals-like approach to the ion-gas thermodynamic description that leads to the manifestation of a sort of Law of Corresponding States.

Finally, the slip correction factor can be cast in the form
\begin{equation}
\label{eq:MC}
%\phi(N,T)= f_0\mathrm{e}^{-T/T_0}\left(
%\frac{N_c}{N}
%\right)^\epsilon \mathrm{e}^{\left(-0.5/K_n^{1.8}+T_c/T\right)}
\phi(N,T)=A\left(\frac{N_c}{N}\right)^\epsilon
\end{equation}
in which \(N_c\approx 80.8 \times 10^{26}\,\)m\(^{-3}\) is the critical density of Argon, \(A\approx 1/4\) and \(\epsilon\approx 1.1\).
\section{Comparison of experimental outcome and theoretical prediction\label{sect:expthe}}
\begin{figure}[h!]
\centering
\includegraphics[width=\columnwidth]{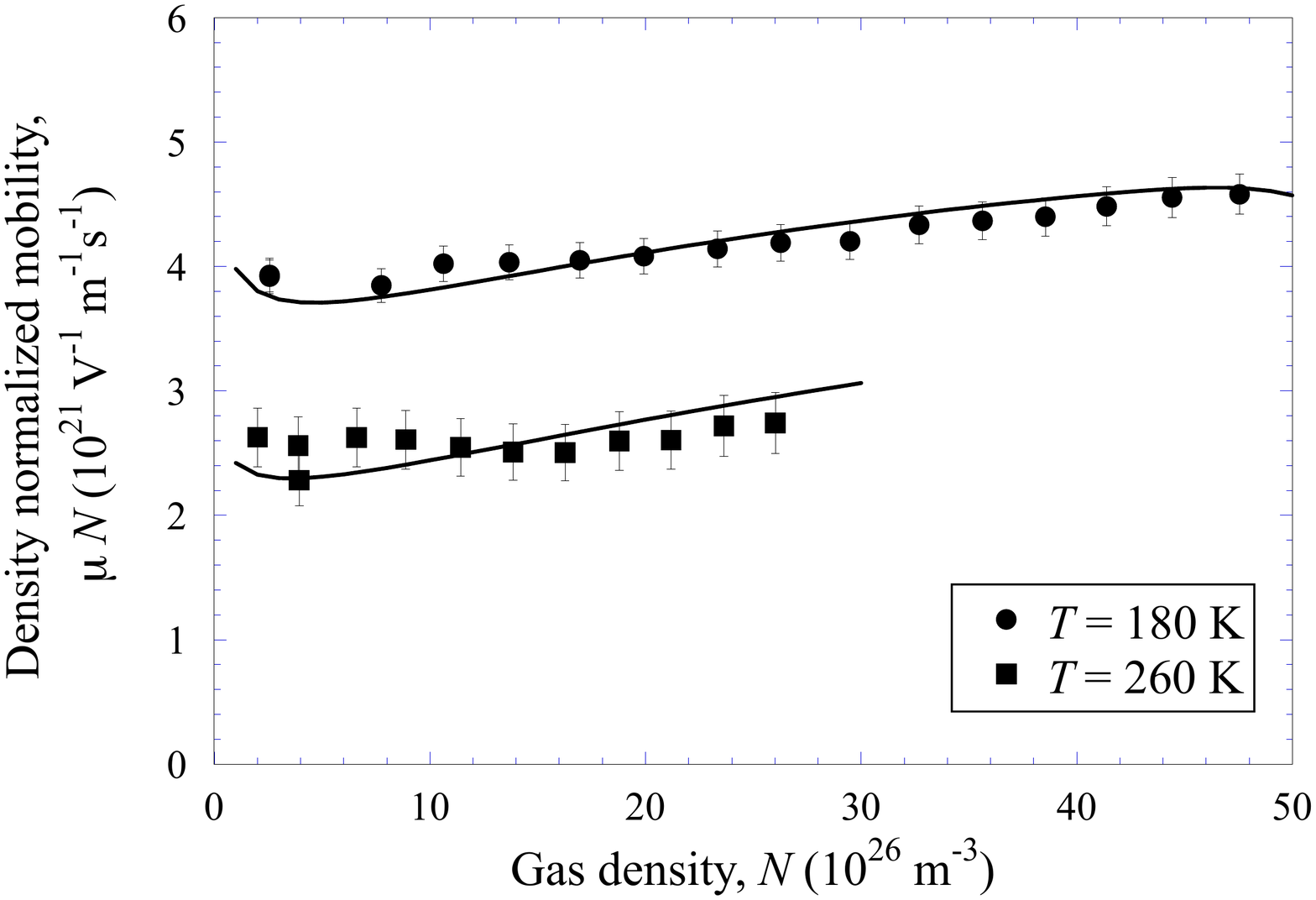}
\caption{\small Comparison of the FV model   results with the experimental outcome.
Symbols: experimental data at \(T=180\,\)K (dots) and \(T=260\,\)K (squares). Line: FV model predictions.
\label{fig:O2-ArmuNvsNT180KeT260K}}
\end{figure}

In Fig.~\ref{fig:O2-ArmuNvsNT180KeT260K} we compare the predictions of the FV model with the experimental data for \(T=180\,\)K and \(T=260\,\)K. Similar results are obtained for all other investigated isotherms.
We note that the values of the free parameters appearing in~(\ref{eq:PAI}), in (\ref{eq:RsuR0}), and in (\ref{eq:MC}) have been determined once for all by seeking agreement between the \(T=180\,\)K experimental data and the model and retain their values through all other isotherms.
As we can see, the agreement between model and experiment is satisfactory. 

We note that the value of the parameter \(\epsilon\) is slightly larger than 1. This value would be problematic in the limit of zero density. However, the value \(\epsilon\approx 1.1\) is obtained by a global fit of the ratio of the mean free path to the hydrodynamic radius over all the investigated density range. Actually, we found that \(\epsilon\) depends on \(N\) in such a way that \(\epsilon =1\) at low density and \(\epsilon \rightarrow 1.2\) at high density. We believe that this behavior is an artifact of the choice of using the kinetic expression of the mean free path that involves the dynamic viscosity of the gas instead of using the simpler expression \(\ell =1/N\sigma,\)
in which, unfortunately, the scattering cross section \(\sigma\) is not known. We further observe that in the real experiment the true zero-density limit is not even closely approached.  Actually, the lowest density we achieved is of the order of \(1\times 10^{26}\,\)m\(^{-3}\), i.e., roughly 4 times the density of an ideal gas at a pressure of \(1\,\) atm and at a temperature of \(0^\circ\, \)% degrees
 C.%elsius. For this reason, we believe the the small difference of \(\epsilon\) from unity is not very relevant.

The FV model also can predict that the effective hydrodynamic radius bridging 
the hydrodynamic- to the kinetic transport regime over the density crossover region
 is cast in the form
\begin{equation} 
\label{eq:reff}
R_\mathrm{eff} =\frac{R}{1+\phi}
\end{equation}
which has to be compared to its experimental determination 
\(R_\mathrm{eff}= e/6\pi\eta \mu \).
\begin{figure}[b!]
\centering
\includegraphics[width=\columnwidth]{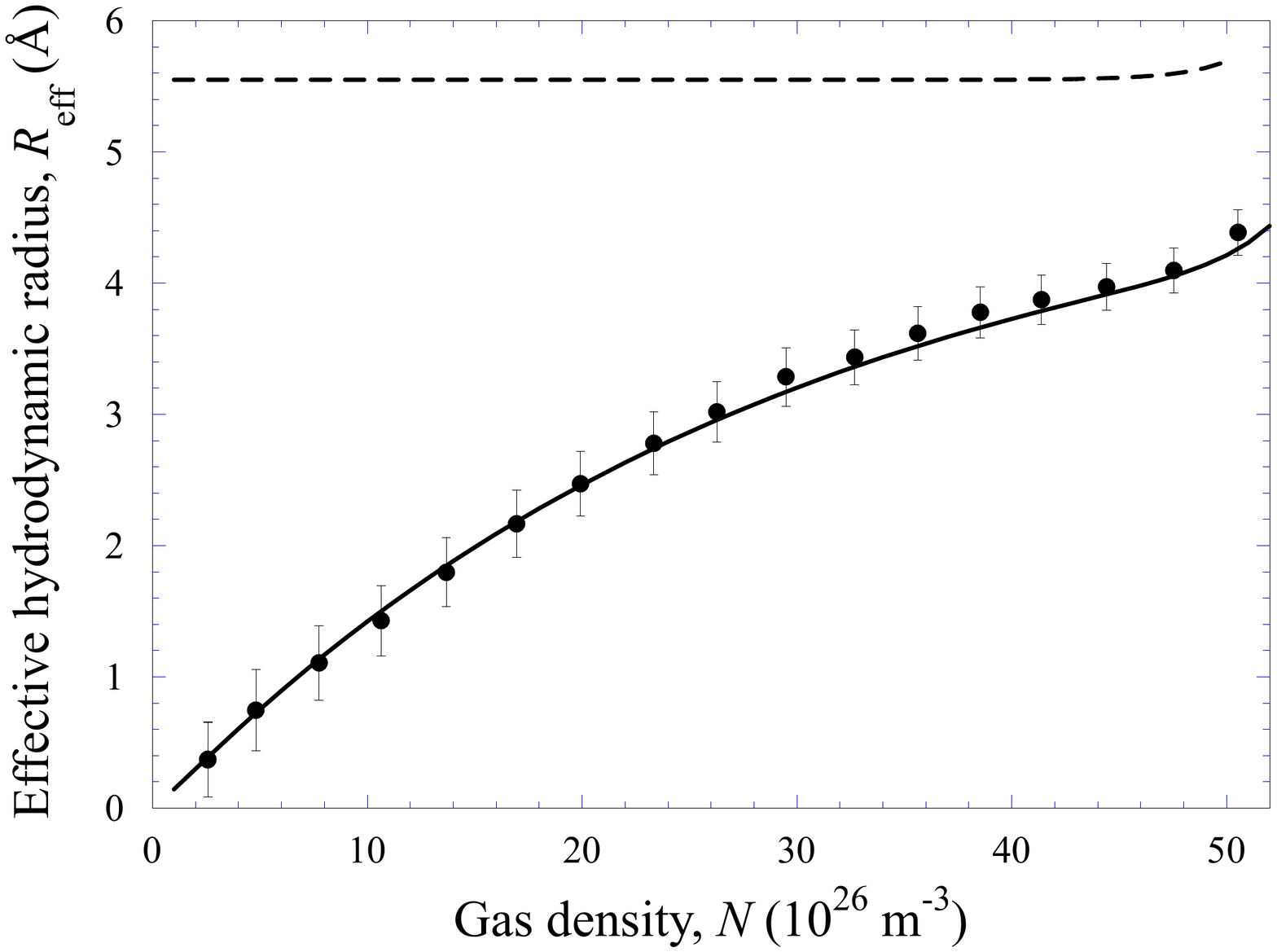}
\caption{\small Density dependence of the effective hydrodynamic radius \(R_\mathrm{eff}\) for \(T=180\,\)K. Points: experimental results. Solid line: FV model prediction. Dashed line: pure hydrodynamic radius.
\label{fig:O2-ReffO2mArT180K}}
\end{figure}
The comparison of the two determinations of the effective radius is shown in Fig.~\ref{fig:O2-ReffO2mArT180K}. We only report the results for \(T=180\,\)K because its investigated density range is the broadest among all isotherms. Moreover, the resulting temperature dependence of \(R_\mathrm{eff}\) is so weak that no further insight is gained by plotting the results for all remaining temperatures.

On one hand, we note that the pure hydrodynamic radius \(R\) is almost constant throughout the whole investigated density range. This fact implies that the experiment has been carried out in the crossover density region but on the low-density side of it. Consequently, the pure Stokes formula would certainly fail at describing the mobility data. 

On the other hand, we see that the agreement between the model prediction and the experimental determination of the effective radius is quite good. We observe that \(R_\mathrm{eff}\) linearly depends on \(N\) at very low density. This means that in this limit \(\mu N\) turns out to be independent of the density as required by the kinetic theory.  As the density is increased, the effective radius increases with decreasing slope and tends to a saturation value, thereby approaching the pure hydrodynamic radius and, hence, the hydrodynamic regime.

We can conclude that in the explored temperature and density ranges, in particular at temperatures quite higher than the critical one, the most important factor to get agreement with the experimental data is to implement the results of the free volume model with a suitable form of the slip correction factor. The same conclusions apply to all investigated isotherms.

\begin{figure}[b!]
\centering
\includegraphics[width=\columnwidth]{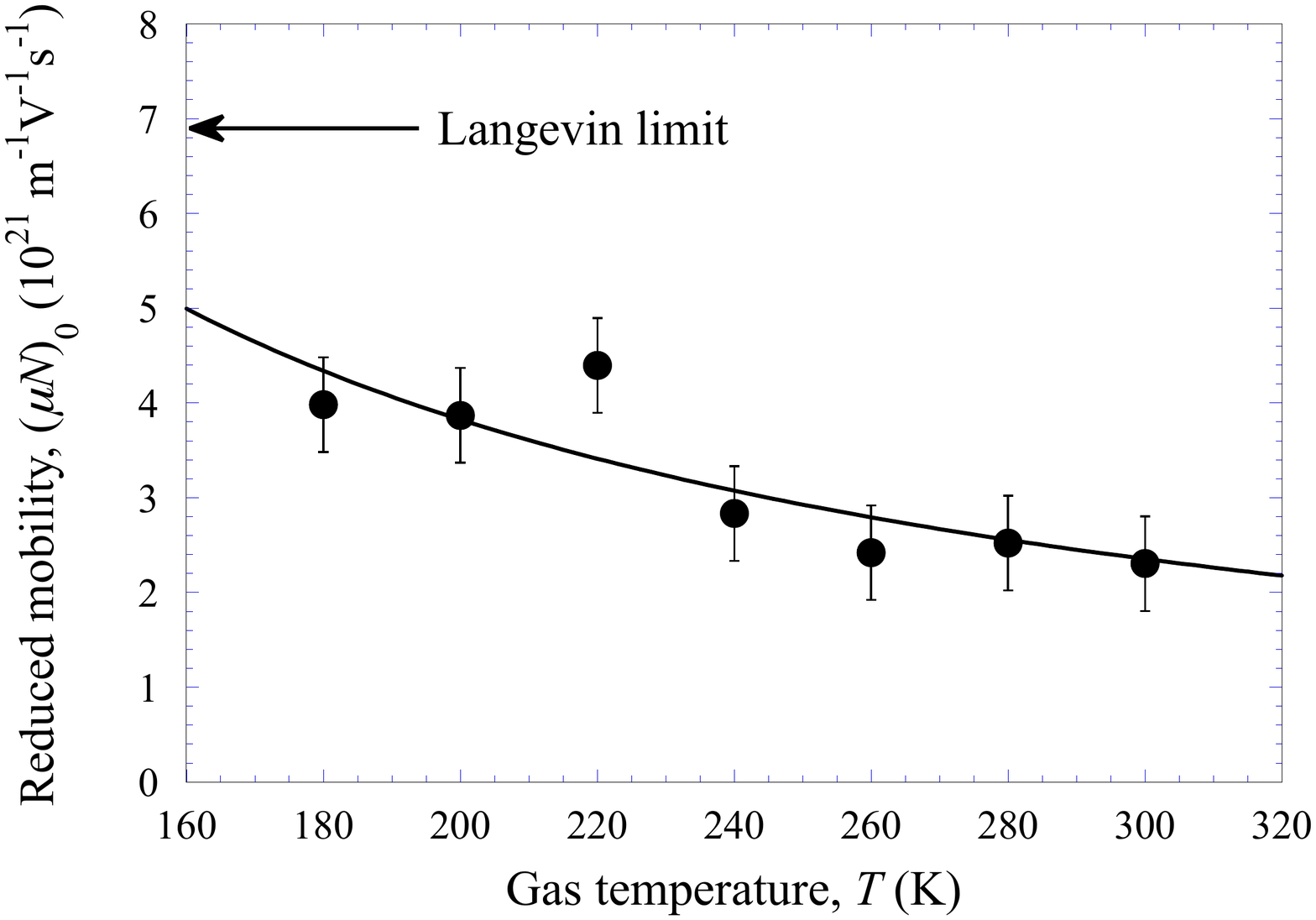}
\caption{\small Temperature dependence of the reduced mobility \(\left(\mu N\right)_0\). Points: experimental results. Line: FV model prediction. The arrow indicates the 
zero-density limit of the mobility of  thermal ions known as {\em polarization} (or, {\em Langevin}) limit.\label{fig:mu0n0O2-ArvsT}}
\end{figure}

Although the model has been developed in such a way to extend the Stokes formula to relatively low densities, nonetheless it also predicts quite correctly the temperature dependence of the zero-density limit of the density-normalized mobility, \(\left(\mu N\right)_0\), defined as 
\begin{equation}
\label{eq:mred}
 \left(\mu N\right)_0=\lim\limits_{N\rightarrow 0} \mu N
\end{equation}
We show in Fig.~\ref{fig:mu0n0O2-ArvsT} the temperature dependence of the experimentally determined \(\left(\mu N\right)_0\) and the prediction of the FV model.
In spite that the accuracy of  \(\left(\mu N\right)_0 \) is relatively poor because it is obtained by extrapolating the available experimental data to \(N\rightarrow 0,\) nonetheless it  appears that the FV model gives a satisfactory description of the experimental outcome. We must stress the fact that, although we reached the lowest possible densities in the experiment, nonetheless we are by far distant from the pure kinetic regime. Actually, in the limit \(N\rightarrow 0\), \(\left(\mu N\right)_0\) should be be given by the so-called polarization (or, Langevin) limit~\cite{heiche1970}
\begin{equation}
\label{eq:lang}
\left(\mu N\right)_0 =
\frac{4.81\times 10^{-4}}{\sqrt{m_r\alpha_s}}\approx 6.9\times 10^{21}
\,\mbox{(V m s)}^{-1}
\end{equation}
in which \(\alpha_s= 11.08\, a_0^3\) is the atomic polarizability of Argon, \(a_0\) is the Bohr radius, and  \(m_r\approx 2.95\times 10^{-25}\,\)kg is the O\(_2^-\)-Ar reduced mass. 
This picture once more confirms the efficacy of the FV model at treating the transition transport regime.
\IEEEpubidadjcol

\section{Conclusion}
In this paper we have reported on new measurements of the O\(_2^-\) ion mobility in dense Argon gas in quite extended temperature and density ranges and on their theoretical rationalization with the free volume model. The measurements are carried out for densities that are too large for the pure kinetic theory to apply and too low for the hydrodynamic theory to be valid. The measurements thus explore a crossover region between the two limiting transport regimes.

We have shown, by means of Molecular Dynamics simulations, that the structure of the fluid surrounding the ions does determine their transport behavior. 

The free volume model that treats the ions as a low concentration solute in a solvent is just aimed at computing the volume associated with each ion that is available for its motion by introducing a van der Waals-like equation of state for the solute.  The linear size of this volume is assumed to be the ion hydrodynamic radius to appear in the hydrodynamical Stokes formula.
 
The model, although relying on some adjustable parameters, nonetheless shows some universal features. First, the hydrodynamic radius turns out to be a function of the free volume. Moreover, the expressions for the internal pressure \(\Pi\) and of the volume of a reference thermodynamical state \(V_0\) turn out to be common to all investigated systems and are thus universal. We believe that this universality stems from a sort of Law of Corresponding States following the van der Waals-like approach. Finally, the model parameter, adjusted to fit one specific isotherm, keep their values though all remaining isotherms.

To bridge the high- to the low-density regions the Stokes formula for the mobility is supplemented with the introduction of a suitable slip correction factor in the wake of the work of Millikan and Cunningham.  By so doing, the FV model quite well describes the mobility in dense Argon gas in the crossover region. Similar good results were also obtained for the mobility of O\(_2^-\) ions in dense Neon gas, of electron bubbles in normal liquid, and superfluid Helium and in several more systems, including cations in different liquids.

Owing to the efficacy of the model, we plan in the next future to apply it to the O\(_2^-\) mobility in dense Argon gas very near to the critical temperature where the large gas compressibility presents a harsh challenge for testing the versatility of the free volume model.

%\section*{Acknowledgments}
%This work benefited from the support of the project ZEROUATE under Grant %ANR-19-CE24-0013 operated by the French National Research Agency (ANR).

\bibliographystyle{IEEEtran}

\begin{thebibliography}{10}
\providecommand{\url}[1]{#1}
\csname url@samestyle\endcsname
\providecommand{\newblock}{\relax}
\providecommand{\bibinfo}[2]{#2}
\providecommand{\BIBentrySTDinterwordspacing}{\spaceskip=0pt\relax}
\providecommand{\BIBentryALTinterwordstretchfactor}{4}
\providecommand{\BIBentryALTinterwordspacing}{\spaceskip=\fontdimen2\font plus
\BIBentryALTinterwordstretchfactor\fontdimen3\font minus
  \fontdimen4\font\relax}
\providecommand{\BIBforeignlanguage}[2]{{%
\expandafter\ifx\csname l@#1\endcsname\relax
\typeout{** WARNING: IEEEtran.bst: No hyphenation pattern has been}%
\typeout{** loaded for the language `#1'. Using the pattern for}%
\typeout{** the default language instead.}%
\else
\language=\csname l@#1\endcsname
\fi
#2}}
\providecommand{\BIBdecl}{\relax}
\BIBdecl

\bibitem{lopez2005}
I.~M. Lopez and V.~Chepel, \emph{{Electronic Excitations in Liquefied Rare
  Gases}}.\hskip 1em plus 0.5em minus 0.4em\relax Stevenson Ranch, CA (USA):
  American Scientific Publishers, 2005, ch. {Rare Gas Liquid Detectors}, pp.
  331--388.

\bibitem{Bruggeman2016b}
P.~J. Bruggeman and {\em et al.}, ``{Plasma-liquid interactions: A review and
  roadmap},'' \emph{Plasma Sources Sci. Technol.}, vol.~25, p. 053002, 2016.

\bibitem{mason2001}
P.~Hughes and N.~Mason, \emph{{Introduction to Environmental Physics: Planet
  Earth, Life and Climate}}.\hskip 1em plus 0.5em minus 0.4em\relax Boca Raton:
  CRC Press, 2001.

\bibitem{viehland1976}
H.~W. Ellis, R.~Y. Pai, E.~W. McDaniel, E.~A. Mason, and L.~A. Viehland,
  ``{Transport properties of Gaseous Ions over a Wide Energy Range},''
  \emph{At. Data and Nucl. Data Tables}, vol.~17, pp. 177--210, 1976.

\bibitem{Viehland1995}
L.~A. Viehland and C.~C. Kirkpatrick, ``{Relating ion/neutral reaction rate
  coefficients and cross-sections by accessing a database for ion transport
  properties},'' \emph{Int. J. Mass Spectrom. Ion Process.}, vol. 149-150,
  no.~C, pp. 555--571, 1995.

\bibitem{viehland1975}
L.~A. Viehland and E.~A. Mason, ``{Tables of transport collision integrals for
  \((n,6,4)\) ion-neutral potentials},'' \emph{At. Data and Nucl. Data Tables},
  vol.~16, pp. 495--514, 1975.

\bibitem{viehland1975b}
------, ``{Gaseous ion mobility in electric fields of arbitrary strength},''
  \emph{Ann. Phys.}, vol.~91, pp. 499--533, 1975.

\bibitem{maitland}
G.~C. Maitland, M.~Rigby, and W.~A. Wakeham, \emph{{Intermolecular Forces.
  Their Origin adn Determination}}.\hskip 1em plus 0.5em minus 0.4em\relax
  Oxford: Clarendon Press, 1981.

\bibitem{byron}
R.~Byron~Bird, W.~E. Stewart, and E.~M. Lightfoot, \emph{{Transport
  phenomena}}.\hskip 1em plus 0.5em minus 0.4em\relax New York: Wiley, 1960.

\bibitem{mason1988}
E.~A. Mason and E.~W. McDaniel, \emph{{Transport Properties of Ions in
  Gases}}.\hskip 1em plus 0.5em minus 0.4em\relax New York: Wiley, 1988.

\bibitem{Borghesani1993}
A.~F. Borghesani, D.~Neri, and M.~Santini, ``{Low-temperature O\(_2^-\)
  mobility in high-density neon gas},'' \emph{Phys. Rev. E}, vol.~48, pp.
  1379--1389, 1993.

\bibitem{Borghesani1995}
A.~F. Borghesani, F.~Chiminello, D.~Neri, and M.~Santini, ``{O\(^-_2\) ion
  mobility in compressed He and Ne Gas},'' \emph{Int. J. Thermophys.}, vol.~16,
  pp. 1235--1244, 1995.

\bibitem{Borghesani1997}
A.~F. Borghesani, D.~Neri, and A.~Barbarotto, ``{Mobility of O\(_2^-\) ions in
  near critical Ar gas},'' \emph{Chem. Phys. Lett.}, vol. 267, pp. 116--122,
  1997.

\bibitem{Bradbury1933}
N.~E. Bradbury, ``{Electron attachment and negative ion formation in oxygen and
  oxygen mixtures},'' \emph{Phys. Rev.}, vol.~44, pp. 883--890, 1933.

\bibitem{christophourou1984a}
L.~G. Christophorou, D.~L. McCorkle, and A.~A. Christodoulides,
  \emph{{Electron-Molecule Interactions and Their Applications}}.\hskip 1em
  plus 0.5em minus 0.4em\relax Orlando: Academic Press, 1984, vol.~I, ch.
  {Electron Attachment Processes}.

\bibitem{Volykhin1995}
K.~F. Volykhin and A.~G. Khrapak, ``{Structure and mobility of negative ions in
  dense gases and nonpolar liquids},'' \emph{JETP}, vol.~81, pp. 901--908,
  1995.

\bibitem{Khrapak1995}
A.~G. Khrapak, W.~F. Schmidt, and K.~F. Volykhin, ``{Structure of O\(_2^-\) in
  dense helium gas},'' \emph{Phys. Rev. E}, vol.~51, pp. 4804--4806, 1995.

\bibitem{Volykhin1999}
K.~F. Volykhin and A.~G. Khrapak, ``{Structure and mobility of negative ions in
  dense gases and nonpolar liquids},'' \emph{JETP}, vol.~81, p. 901, 1999.

\bibitem{Schmidt1999}
W.~F. Schmidt, K.~F. Volykhin, A.~G. Khrapak, and E.~Illenberger, ``{Structure
  and mobility of positive and negative ions in non-polar liquids},'' \emph{J.
  Electrostat.}, vol.~47, pp. 83--95, 1999.

\bibitem{Borghesani1999}
A.~Borghesani, D.~Neri, and A.~Barbarotto, ``{Critical behavior of O2 - ions in
  Argon gas},'' \emph{Int. J. Thermophys.}, vol.~20, pp. 899--909, 1999.

\bibitem{Borghesani2019}
\BIBentryALTinterwordspacing
A.~F. Borghesani and F.~Aitken, ``{A thermodynamic model for O\(_2^-\) mobility
  in neon gas over broad density and temperature ranges},'' \emph{Plasma
  Sources Science and Technology}, vol.~28, p. 03LT01, 2019. [Online].
  Available: \url{https://doi.org/10.1088/1361-6595/ab071a}
\BIBentrySTDinterwordspacing

\bibitem{Borghesani2020}
------, ``O\(_2^-\) ion mobility in dense ne gas: The free volume model,''
  \emph{IEEE Trans. Dielectr. Electr. Insul.}, vol.~27, pp. 757--763, 2020.

\bibitem{Aitken2011}
F.~Aitken, Z.-L. Li, N.~Bonifaci, A.~Denat, and K.~von Haeften, ``{Electron
  mobility in liquid and supercritical helium measured using corona discharges:
  a new semi-empirical model for cavity formation.}'' \emph{Phys. Chem. Chem.
  Phys.}, vol.~13, pp. 719--724, 2011.

\bibitem{Aitken2011a}
F.~Aitken, N.~Bonifaci, A.~Denat, and K.~{Von Haeften}, ``{A macroscopic
  approach to determine electron mobilities in low-density helium},'' \emph{J.
  Low Temp. Phys.}, vol. 162, pp. 702--709, 2011.

\bibitem{Aitken2017}
F.~Aitken, F.~Volino, L.~Mendoza-Luna, K.~Haeften, and J.~Eloranta, ``{A
  thermodynamic model to predict electron mobility in superfluid helium},''
  \emph{Phys. Chem. Chem. Phys.}, vol.~19, pp. 15\,821--15\,832, 2017.

\bibitem{Cunningham1910}
E.~Cunningham, ``{On the Velocity of Steady Fall of Spherical Particles through
  Fluid Medium Author ( s ): E . Cunningham Source : Proceedings of the Royal
  Society of London . Series A , Containing Papers of a Published by : Royal
  Society Stable URL : http://www.jstor.org/},'' vol.~83, pp. 357--365, 1910.

\bibitem{Millikan1910}
R.~A. Millikan, ``{The Isolation of an Ion, A Precision Measurement of Its
  Charge, and the Correction of Stokes},'' \emph{Science}, vol.~32, pp.
  436--448, 1910.

\bibitem{Tyndall1938}
A.~M. Tyndall, \emph{The mobility of positive ions in gases}.\hskip 1em plus
  0.5em minus 0.4em\relax Cambridge: The University Press, 1938.

\bibitem{wagner}
C.~Tegeler, R.~Span, and W.~Wagner, ``{A New Equation of State for Argon
  Covering the Fluid Region for Temperatures From the Melting Line to 700 K at
  Pressures up to 1000 MPa},'' \emph{J. Phys. Chem. Ref. Data}, vol.~28, pp.
  779--850, 1999.

\bibitem{Borghesani1990a}
A.~F. Borghesani and M.~Santini, ``{Electron swarm experiments in fluids-
  signal waveform analysis},'' \emph{Meas. Sci. Tecnnol.}, vol.~1, pp. 939 --
  947, 1990.

\bibitem{Borghesani2018}
A.~F. Borghesani and F.~Aitken, ``{Molecular dynamics simulations of the
  O\(_2^-\) ion mobility in dense Ne gas at low temperature: Influence of the
  repulsive part of the ion-neutral interaction potential},'' \emph{IEEE Trans.
  Dielectr. Electr. Insul.}, vol.~25, pp. 1992--1998, 2018.

\bibitem{Aitken2016}
F.~Aitken, N.~Bonifaci, K.~{Von Haeften}, and J.~Eloranta, ``{Theoretical
  modeling of electron mobility in superfluid 4He},'' \emph{J. Chem. Phys.},
  vol. 145, no.~4, p. 044105, 2016.

\bibitem{Aitken2015}
F.~Aitken, N.~Bonifaci, L.~G. Mendoza-Luna, and K.~von Haeften, ``{Modelling
  the mobility of positive ion clusters in normal liquid helium over large
  pressure ranges},'' \emph{Phys. Chem. Chem. Phys.}, vol.~17, pp.
  18\,535--18\,540, 2015.

\bibitem{Reif}
F.~Reif, \emph{{Fundamentals of Statistical and Thermal Physics}},
  18th~ed.\hskip 1em plus 0.5em minus 0.4em\relax Auckland: McGraw-Hill, 1985.

\bibitem{heiche1970}
G.~Heiche and E.~A. Mason, ``{Ion mobilities with charge exchange},'' \emph{J.
  Chem. Phys.}, vol.~53, pp. 4687--4696, 1970.

\end{thebibliography}


%\newpage
%\vspace{7pt}
\section*{Biography Section}
\vspace{-33pt}
\begin{IEEEbiography}[{\includegraphics[width=1in,height=1.25in,clip,keepaspectratio]{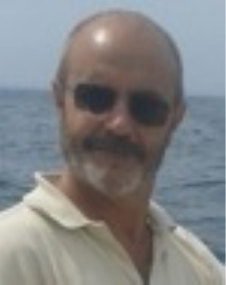}}]{Armando Francesco Borghesani}
Born in Verona (Italy), 1956. %After completing a classics high school  (1974), he obtained %his 
MS in Physics in 1979 with a thesis on  ``Anomalous behavior of the CO2 viscosity at the critical point''. 
From  5/1980 to  11/1981 he joined %SGS-ATES (now 
ST Microelectronics in Milan (Italy)
% ), a semiconductor manufacturer in Milan (Italy) 
as junior researcher in the plasma etching research group.
From 11/1981 to 7/1983 he joined ENI-Ricerche in Milan (Italy) as a researcher in the department for coal slurries pretreatment and conditioning.
From 9/1983 to 10/1992 he was appointed as Assistant Professor in Physics of Matter at the Physics Department of the University of Padua (Padova, Italy).
Since 11/1992 he is Associate Professor of Experimental Physics in the Physics \& Astronomy Department of the same University. He taught courses in Mechanics, Electromagnetism and Optics, Quantum and Solid State Physics, Electronics.
 He investigated critical point phenomena, rheology of coal-based slurries, plasma etching of semiconductors, energetics and dynamics of electrons and ions in dense noble gases and liquids, infrared luminescence of noble gas excimers in high density gaseous environment, cathodoluminescence of rare-earth doped non-linear crystals for developing dark-matter particle detectors, and thermal radiation of hot metal bodies.
He established collaborations with groups at the Max-Planck Institute (Munich, Germany), at the Vrije Universiteit (Amsterdam, The Netherlands), and at CNRS (Grenoble, France).
He is author of more than 100 peer-reviewed papers and of one monography for Oxford University Press entitled ``Ions and Electrons in Liquid Helium'' in addition to textbooks in Quantum Physics (in Italian).
He has contributed, several times as invited speaker, to more than 40 international conferences and schools. He is actively contributing to all International Conferences on Dielectric Liquids (ICDL) since 1990 and was awarded the Hans Tropper Award at ICDL 2022.

\end{IEEEbiography}
\vfill

\vspace{-18pt}

%\bf{If you include a photo:}
\vspace{-33pt}
\begin{IEEEbiography}
[{\includegraphics[width=1in,height=1.25in,clip,keepaspectratio]{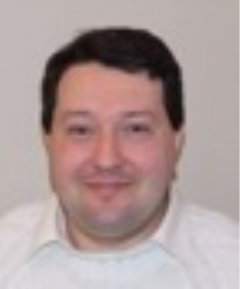}}]{Fr\'ed\'eric Aitken}
(born in Grenoble, France, 1969) graduated in Mechanics from Grenoble Joseph Fourier University and Aeronautical Engineering School in Toulouse. In1994-1995 he was appointed engineer at Institute Max Planck, where he was in charge of the study of cooling of resistive magnetic coils of Polyhelix type. He passed his Ph. D. degree in 1998 and joined the CNRS as a full researcher shortly after. 2010-2014, he was appointed as President of the French Physical Society (SFP-Alpes). His researches are focused on the investigation of fast energy injection in dielectric liquids for characterization of phase change induced by corona-like discharges. Making relations of his own with renowned scientists he became a specialist of non-equilibrium thermodynamic and transport properties in these fluids.
\end{IEEEbiography}

\vfill

\end{document}